\documentclass[12pt]{article}
\usepackage{times}
\usepackage{a4}
\usepackage{latexsym}
\usepackage{amsmath}
\usepackage{amsfonts}
\usepackage{graphicx}
\usepackage{fancyhdr,lastpage}
\newtheorem{theorem}{Theorem}
\newenvironment{proof}[1][Proof]{\noindent\textbf{#1.} }{\ \rule{0.5em}{0.5em}}
\begin{document}
\title{Exactly solvable model for the dynamics of two spin-$\frac{1}{2}$ particles embedded in separate spin star environments}

\author{Yamen Hamdouni\thanks{Suggestions and
corrections welcome} \thanks{Email: {\tt hamdouniyamen@gmail.com}.}} 
 \maketitle



\begin{abstract}
Exact analytical results for the dynamics of two interacting qubits
each of which is embedded in its own spin star bath  are presented.
The time evolution of the concurrence and the purity of the
two-qubit system is investigated for finite and infinite numbers of
environmental spins. The effect of qubit-qubit interactions on the
steady state of the central system is investigated.
\end{abstract}

\section{Introduction}
Exactly solvable models play a very  useful  role in various fields
of physics.  They help  improving our understanding of
 physical processes  and  allow us  gain more insight into
  complicated phenomena that take place in nature~\cite{1}. Needless to
 recall the usefulness  of  exactly solvable models such as
the harmonic oscillator, the nuclear shell model and the Ising
model, to mention but a few. From a practical point of view, exactly
solvable models serve as a very convenient tool for testing the
accuracy of numerical algorithms, often used in the study of
problems that cannot be analytically solved due to the complexity of
the systems under investigation.

In nature, quantum systems are influenced by their surrounding
environment through, in general, complicated coupling interactions,
leading them to lose their coherence~\cite{2}. This refers to as the
decoherence process~\cite{3,4,5}.  Moreover, quantum systems exhibit
properties that do not have classical analogous~\cite{6}. Of great
interest is entanglement, the main ingredient for quantum
teleportation and quantum computation~\cite{7,8,9,10,11,12}. Over
the last years, many proposals have been made for the implementation
of quantum information processing. Solid state systems are very
promising~\cite{13,14} and have been the subject of many
investigations. In particular, decoherence and entanglement of
qubits coupled to spin environments~\cite{15} attracted much
attention~\cite{16,17}.  Thus new exactly solvable models describing
the dynamics of qubits in spin baths are highly welcome. Recently,
the spin star configuration, initially proposed by Bose, has been
extensively investigated~\cite{18,19,20,21,22}. An exact treatment
of the dynamics of two qubits coupled to common spin star bath via
$XY$ interactions is presented in~\cite{23,24}. In this paper we
propose to investigate analytically the dynamics when the two qubits
interact with separate spin star baths.

The paper is organized as follows. In section~\ref{sec2} the model
Hamiltonian is introduced. In section~\ref{sec3} we present a
detailed derivation of the time evolution operator and we
investigate the dynamics of the qubits at finite $N$ for some
particular initial conditions. In section~\ref{sec4} we study the
thermodynamic limit, in which the sizes of the spin environments
become infinite. Section~\ref{sec5} is devoted to the second-order
master equation. We end the paper with a short summary.

\section{Model\label{sec2}}

 The system under study consists of two two-level systems (
    { e.g.,} spin-$\frac{1}{2}$ particles) each of which is
    embedded in its own spin star
environment composed of $N$ spins-$\frac{1}{2}$. The central
particles interact with each other through a Ising interaction; the
corresponding coupling constant is equal to $4\delta$, where the
factor 4 is introduced for later convenience. We shall  assume that
 each qubit  couples to  its environment via Heisenberg $XY$
interaction whose coupling constant is $\alpha$, which is, in turn,
scaled by $N^{1/2}$ in order to ensure good thermodynamic behavior.
The spin baths will be denoted by $B_1$ and $B_2$. The Hamiltonian
for the composite system has the form
\begin{equation}
H=H_0+H_{S_1B_1}+H_{S_2B_2},
\end{equation}
where
\begin{equation}
H_0=4\delta S^{1}_z S^{2}_z,\end{equation} and
\begin{equation}
H_{S_iB_i}=\frac{\alpha}{\sqrt{N}} (S_+^{i}\sum_{k=1}^N S^{ik}_-+
S_-^{i}\sum_{k=1}^N S^{ik}_+), \quad (i=1,2).
\end{equation}
 Here  $\vec{S^1}$ and $\vec{S^2}$ denote the spin operators
 corresponding to the central qubits, whereas $\vec{S^{ik}}$ denotes
the spin operator  corresponding to the $k^{th}$ particle within the
$i^{th}$ environment. Introducing the total spin operators
$\vec{J}=\sum_{k=1}^N\vec{S^{1k}}$ and $\vec{\mathcal
J}=\sum_{k=1}^N\vec{S^{2k}}$ of the environments $B_1$ and $B_2$,
respectively, one can rewrite the full Hamiltonian as
\begin{equation}
H=4\delta
S^1_zS^2_z+\frac{\alpha}{\sqrt{N}}(S^1_+J_-+S^1_-J_++S^2_+\mathcal{J}_-
+S^2_-\mathcal{J}_+)\label{fullh}.\end{equation}

The dynamics of the two-qubit system is fully described by its
density matrix $\rho(t)$ obtained, as usual, by tracing the
time-dependent total density matrix $\rho_{\rm tot}(t)$, describing
the composite system, with respect to the  environmental degrees of
freedom, namely,
\begin{eqnarray}
\rho(t)&=&{\rm tr}_{B_1+B_2}[\rho_{\rm tot}(t)]\nonumber \\
&=& {\rm tr}_{B_1+B_2}\Bigl[\mathbf U(t)\rho_{\rm tot}(0)\mathbf
U^\dag(t)\Bigl],
\end{eqnarray}
where $\mathbf U(t)$ and $\rho_{\rm tot}(0)$ designate the time
evolution operator and the initial total density matrix,
respectively.

At $t=0$ the central qubits are assumed to be uncoupled with the
environments; the latter are assumed to be at infinite temperature.
This means that the initial total density density matrix can be
written as
\begin{equation}
\rho_{\rm tot}(0)=\rho(0)\otimes\frac{\mathbf 1 }{2^N}\otimes
\frac{\mathbf 1 }{2^N}.\end{equation} Here $\rho(0)$ is the initial
density matrix of the two-qubit system, and $\mathbf 1$ is the unit
matrix on the space $\mathbb C^{2\otimes N}$. The former can be
written as $\rho(0)=\sum_{k,\ell,}
\rho_{k\ell}^0|\chi_k\rangle\langle\chi_\ell|$, with
$|\chi_\ell\rangle\in\{|--\rangle,|-+\rangle,|+-\rangle,|++\rangle\}$
for $\ell=\overline{1,4}$. Similarly, we introduce the basis state
vectors $|j,m\rangle$ of $\mathbb C^{2\otimes N}$, such that
$\kappa\le j \le N/2$ ($\kappa=0$ for $N$ even and $\kappa=1/2$ for
$N$ odd), and $-j\le m\le j$. The time-dependent reduced density
matrix can be expressed as
\begin{equation}
\rho(t)=2^{-2N}\sum_{k,\ell}\rho^0_{k\ell}\sum_{j,m}\sum_{r,s}\nu(N,j)\nu(N,r)\langle
j,r,m,s|\mathbf U(t)|\chi_k\rangle\langle \chi_\ell|\mathbf
U^\dag(t)|j,r,m,s\rangle,\end{equation} where
$|j,r,m,s\rangle=|j,m\rangle\otimes|r,s\rangle$, and
$\nu(N,j)=\binom{N}{N/2-j}-\binom{N}{N/2-j-1}$~\cite{25}. Hence, our
task reduces to finding the exact form of the matrix elements of the
time evolution operator ${\mathbf U}(t)=\exp(-iHt)$ ($\hbar=1$).
This will be the subject of the next section.

\section{Derivation of the exact form of the time evolution operator\label{sec3}}

The time evolution operator can be expanded as
\begin{equation} \mathbf
U(t)=\sum\limits_{n=0}^{\infty} \frac{{(-1)^n
t^{2n}}}{(2n)!}(H)^{2n}-i\sum\limits_{n=0}^{\infty}\frac{(-1)^n
t^{2n+1}}{(2n+1)!}( H)^{2n+1} \label{taylor}.
\end{equation}
In order to derive analytical expressions for even and odd powers of
the total Hamiltonian $H$ let us notice that $H_0$ anticommutes with
$H_{S_1B_1}+H_{S_2B_2}$, that is,
\begin{equation}
[H_0,H_{S_1B_1}+H_{S_2B_2}]_+=0.
\end{equation}
This can easily be shown using the following  properties for
spin-$\frac{1}{2}$ operators: $S_z S_\pm=\pm S_\pm$, and  $S_\pm
S_z=\mp S_\pm$. Moreover, it is easily seen that
$H_0^{2n}\equiv\delta^{2n}$, which simply implies that for $n \ge0$,
\begin{eqnarray}
H^{2n}&=&\sum\limits_{\ell=0}^{n}\binom{n}{\ell}(H_{S_1B_1}+H_{S_2B_2})^{2\ell}\delta^{2(n-\ell)}\label{binom}.
\end{eqnarray}

 In the standard
basis of $\mathbb C^2\otimes \mathbb C^2$, it can be shown that
powers of $H_{S_1B_1}$ and $H_{S_2B_2}$ are given by
\begin{eqnarray}
H_{S_1B_1}^{2k}&=&\Bigl(\frac{\alpha}{\sqrt{N}}\Bigl)^{2k}\begin{pmatrix}(J_+J_-)^k&&0&&0&&0\\0&&(J_+J_-)^k&&0&&0\\
0&&0&&(J_-J_+)^k&&0\\0&&0&&0&&(J_-J_+)^k\end{pmatrix},\\
H_{S_1B_1}^{2k+1}&=&\Bigl(\frac{\alpha}{\sqrt{N}}\Bigl)^{2k+1}\begin{pmatrix}0&&0&&J_+(J_-J_+)^k&&0\\
0&&0&&0&&J_+(J_-J_+)^k\\J_-(J_+J_-)^k&&0&&0&&0\\0&&J_-(J_+J_-)^k&&0&&0\end{pmatrix},\\
H_{S_2B_2}^{2k}&=&\Bigl(\frac{\alpha}{\sqrt{N}}\Bigl)^{2k}\begin{pmatrix}(\mathcal J_+\mathcal J_-)^k&&0&&0&&0\\0&&(\mathcal J_-\mathcal J_+)^k&&0&&0\\
0&&0&&(\mathcal J_+\mathcal J_-)^k&&0\\0&&0&&0&&(\mathcal J_-\mathcal J_+)^k\end{pmatrix},\\
H_{S_2B_2}^{2k+1}&=&\Bigl(\frac{\alpha}{\sqrt{N}}\Bigl)^{2k+1}\begin{pmatrix}0&&\mathcal J_+(\mathcal J_-\mathcal J_+)^k&&0&&0\\
\mathcal J_-(\mathcal J_+\mathcal J_-)^k&&0&&0&&0\\0&&0&&0&&\mathcal
J_+(\mathcal J_-\mathcal J_+)^k\\0&&0&& \mathcal J_-(\mathcal
J_+\mathcal J_-)^k&&0\end{pmatrix}.
\end{eqnarray}
It follows that
\begin{eqnarray}
(H_{S_1B_1}+
H_{S_2B_2})^{2\ell}&=&\sum\limits_{k=0}^{\ell}\binom{2\ell}{2k}H_{S_1B_1}^{2k}H_{S_2B_2}^{2(\ell-k)}+
\sum\limits_{k=0}^{\ell-1}\binom{2\ell}{2k+1}H_{S_1B_1}^{2k+1}H_{S_2B_2}^{2(\ell-k)-1}\nonumber
\\
&=&\Bigl(\frac{\alpha}{\sqrt{N}}\Big)^{2\ell}\Biggl[\sum\limits_{k=0}^{\ell}\binom{2\ell}{2k}
D_{\ell k} +\sum\limits_{k=0}^{\ell-1}\binom{2\ell}{2k+1}L_{\ell
k}\Biggl].
\end{eqnarray}
where
\begin{eqnarray}
D_{\ell k}=\mathit{diag}\Bigl[
 (J_+J_-)^k(\mathcal J_+\mathcal
J_-)^{\ell-k}, (J_+J_-)^k(\mathcal J_-\mathcal
J_+)^{\ell-k},\nonumber \\ (J_-J_+)^k(\mathcal J_+  \mathcal
J_-)^{\ell-k},(J_-J_+)^k(\mathcal J_-\mathcal
J_+)^{\ell-k}\Bigl]\end{eqnarray} and
\begin{eqnarray}
L_{\ell k}= \mathit{antidiag}\Bigl[ J_+\mathcal J_+
(J_-J_+)^k(\mathcal J_-\mathcal J_+)^{\ell-k-1}, J_+\mathcal J_-
(J_-J_+)^k(\mathcal J_+\mathcal J_-)^{\ell-k-1},\nonumber \\
J_-\mathcal J_+(J_+J_-)^k(\mathcal J_-  \mathcal
J_+)^{\ell-k-1},J_-\mathcal J_+ (J_+J_-)^k(\mathcal J_+\mathcal
J_-)^{\ell-k-1}\Bigl].\end{eqnarray}
 Using the fact that
\begin{eqnarray} \sum\limits_{k=0}^{\ell}
\binom{2\ell}{2k} x^k y^{\ell-k}&=&\frac{1}{2} \Bigl[( \sqrt{x} +
\sqrt{y})^{2\ell}
+ (\sqrt{x} - \sqrt{y})^{2 \ell}\Bigl],\\
 \sum\limits_{k=0}^{\ell-1}
\binom{2\ell}{2k+1} x^k y^{\ell-k-1}&=& \frac{1}{2\sqrt{x y}}\Bigl[(
\sqrt{x} + \sqrt{y})^{2 \ell} - (\sqrt{x} - \sqrt{y})^{2
\ell}\Bigl],\end{eqnarray} one obtains
\begin{eqnarray}
&&(H_{S_1B_1}+H_{S_2B_2})^{2\ell}=\Bigl(\frac{\alpha}{\sqrt{N}}\Bigl)^{2\ell}\nonumber
\\&& \times \begin{pmatrix}F_1^+&&0&&0&&J_+\mathcal J_+
\frac{F_4^-}{\sqrt{J_-J_+\mathcal J_-\mathcal
J_+}}\\0&&F_2^+&&J_+\mathcal J_- \frac{F_3^-}{\sqrt{J_-J_+\mathcal
J_+\mathcal J_-}}&&0\\0&&J_-\mathcal J_+
\frac{F_2^-}{\sqrt{J_+J_-\mathcal J_-\mathcal
J_+}}&&F_3^+&&0\\J_-\mathcal J_- \frac{F_1^-}{\sqrt{J_+J_-\mathcal
J_+\mathcal
J_-}}&&0&&0&&F^+_4\end{pmatrix},\label{powereven}\end{eqnarray}
where
\begin{eqnarray}
F^\pm_1=\frac{1}{2}\Bigl[\Bigl(\sqrt{J_+J_-}+\sqrt{\mathcal
J_+\mathcal J_-}\Bigl)^{2\ell}\pm\Bigl(\sqrt{J_+J_-}-\sqrt{\mathcal
J_+\mathcal J_-}\Bigl)^{2\ell}\Bigl],\\
F^\pm_2=\frac{1}{2}\Bigl[\Bigl(\sqrt{J_+J_-}+\sqrt{\mathcal
J_-\mathcal J_+}\Bigl)^{2\ell}\pm\Bigl(\sqrt{J_+J_-}-\sqrt{\mathcal
J_-\mathcal J_+}\Bigl)^{2\ell}\Bigl],\\
F^\pm_3=\frac{1}{2}\Bigl[\Bigl(\sqrt{J_-J_+}+\sqrt{\mathcal
J_+\mathcal J_-}\Bigl)^{2\ell}\pm\Bigl(\sqrt{J_-J_+}-\sqrt{\mathcal
J_+\mathcal J_-}\Bigl)^{2\ell}\Bigl],\\
F^\pm_4=\frac{1}{2}\Bigl[\Bigl(\sqrt{J_-J_+}+\sqrt{\mathcal
J_-\mathcal J_+}\Bigl)^{2\ell}\pm\Bigl(\sqrt{J_-J_+}-\sqrt{\mathcal
J_-\mathcal J_+}\Bigl)^{2\ell}\Bigl].
\end{eqnarray}
Inserting equation~(\ref{powereven}) into equation~(\ref{binom}),
yields
\begin{eqnarray}
&& H^{2n}=\frac{1}{2}\nonumber\\
 &&\times \begin{pmatrix}(\mathcal M_1^+)^n+(\mathcal M_1^-)^n &&0&&0&&J_+\mathcal
J_+ \frac{(\mathcal M_4^+)^n-(\mathcal
M_4^-)^n}{\sqrt{J_-J_+\mathcal J_-\mathcal J_+}}\\0&&(\mathcal
M_2^+)^n+(\mathcal M_2^-)^n&&J_+\mathcal J_- \frac{(\mathcal
M_3^+)^n-(\mathcal M_3^-)^n}{\sqrt{J_-J_+\mathcal J_+\mathcal
J_-}}&&0\\0&&J_-\mathcal J_+ \frac{(\mathcal M_2^+)^n-(\mathcal
M_2^-)^n}{\sqrt{J_+J_-\mathcal J_-\mathcal J_+}}&&(\mathcal
M_3^+)^n+(\mathcal M^-_3)^n&&0\\J_-\mathcal J_- \frac{(\mathcal
M_1^+)^n-(\mathcal M_1^-)^n}{\sqrt{J_+J_-\mathcal J_+\mathcal
J_-}}&&0&&0&&(\mathcal M^+_4)^n+(\mathcal
M^-_4)^n\end{pmatrix},\label{power1}\end{eqnarray} where
\begin{eqnarray}
\mathcal
M_1^\pm=\delta^2+\frac{\alpha^2}{N}\Bigl(\sqrt{J_+J_-}\pm\sqrt{\mathcal
J_+\mathcal J_-}\Bigl)^2,\\
\mathcal
M_2^\pm=\delta^2+\frac{\alpha^2}{N}\Bigl(\sqrt{J_+J_-}\pm\sqrt{\mathcal
J_-\mathcal J_+}\Bigl)^2,\\
\mathcal
M_3^\pm=\delta^2+\frac{\alpha^2}{N}\Bigl(\sqrt{J_-J_+}\pm\sqrt{\mathcal
J_+\mathcal J_-}\Bigl)^2,\\
\mathcal
M_4^\pm=\delta^2+\frac{\alpha^2}{N}\Bigl(\sqrt{J_-J_+}\pm\sqrt{\mathcal
J_-\mathcal J_+}\Bigl)^2.\end{eqnarray} The above operators satisfy
\begin{eqnarray}
&&M_{1,2}^\pm J_+=J_+M_{3,4}^\pm, \qquad M_{1,2}^\pm \mathcal
J_+=\mathcal J_+M_{3,4}^\pm, \\ && M_1^\pm J_+\mathcal
J_+=J_+\mathcal J_+ M_4^\pm, \qquad  M_2^\pm J_+\mathcal
J_-=J_+\mathcal J_-M_3^\pm.
\end{eqnarray}

Furthermore, one can show  that the matrix elements of $H^{2n+1}$
are given by
\begin{align}
(H^{2n+1})_{11}&=\frac{1}{2}\delta[(\mathcal M_1^+)^n+(\mathcal M_1^-)^n],\label{powerodd1} \\
(H^{2n+1})_{12}&=\mathcal J_+\frac{\alpha}{2\sqrt{N \mathcal
J_-\mathcal J_+}}[(\sqrt{\mathcal J_-\mathcal
J_+}+\sqrt{J_+J_-})(\mathcal M_2^+)^n  \\&+(\sqrt{\mathcal
J_-\mathcal J_+}-\sqrt{J_+J_-})(\mathcal M_2^-)^n], \\
(H^{2n+1})_{13}&=\ J_+\frac{\alpha}{2\sqrt{N  J_-
J_+}}[(\sqrt{\mathcal J_+\mathcal J_-}+\sqrt{J_-J_+})(\mathcal
M_3^+)^n  \\&+(\sqrt{ J_-J_+}-\sqrt{\mathcal J_+\mathcal
J_-})(\mathcal M_3^-)^n], \\
(U^{2n+1})_{14}&=(\delta/2)  J_+\mathcal J_+ \frac{(\mathcal
M_4^+)^n-(\mathcal M_4^-)^n}{\sqrt{J_-J_+\mathcal J_-\mathcal J_+}}
,\end{align}
\begin{eqnarray}
 (H^{2n+1})_{21}&=&\mathcal
J_-\frac{\alpha}{2\sqrt{N \mathcal J_+\mathcal J_-}}[(\sqrt{\mathcal
J_+\mathcal J_-}+\sqrt{J_+J_-})(\mathcal M_1^+)^n
\\&+&(\sqrt{\mathcal
J_+\mathcal J_-}-\sqrt{J_+J_-})(\mathcal M_1^-)^n], \\
(H^{2n+1})_{22}&=&-\frac{1}{2}\delta[(\mathcal M_2^+)^n+(\mathcal M_2^-)^n], \\
(H^{2n+1})_{23}&=&-(\delta/2) J_+\mathcal J_- \frac{(\mathcal
M_3^+)^n-(\mathcal M_3^-)^n}{\sqrt{J_-J_+\mathcal J_+\mathcal
J_-}}, \\
(H^{2n+1})_{24}&=&J_+\frac{\alpha/2}{\sqrt{N J_-
J_+}}[(\sqrt{\mathcal J_-\mathcal J_+}+\sqrt{ J_-
 J_+}) (\mathcal M_4^+)^n  \\&+&(\sqrt{J_-  J_+}-\sqrt{\mathcal J_-\mathcal
J_+})(\mathcal M_4^-)^{n}], \end{eqnarray}
\begin{eqnarray} (H^{2n+1})_{31}&=&
J_-\frac{\alpha/2}{\sqrt{N J_+J_-}}[(\sqrt{J_+J_-}+\sqrt{\mathcal
J_+ \mathcal J_-})( \mathcal M_1^+)^n
\\&+&(\sqrt{J_+J_-}-\sqrt{\mathcal J_+ \mathcal
J_-})(\mathcal M_1^-)^n], \\
 (H^{2n+1})_{32}&=&-(\delta/2)J_-\mathcal J_+\frac{(\mathcal M_2^+)^n-(\mathcal M_1^-)^n}{\sqrt{J_+J_-\mathcal
 J_-\mathcal
J_+}}, \\
(H^{2n+1})_{33}&=&-\frac{1}{2}\delta[(\mathcal M_3^+)^n+(\mathcal M_3^-)^n], \\
(H^{2n+1})_{34}&=&\mathcal  J_+\frac{\alpha/2}{\sqrt{N \mathcal
J_-\mathcal J_+}}[(\sqrt{\mathcal J_-\mathcal J_+}+\sqrt{J_-
J_+})(\mathcal M_4^+)^n \\&+&(\sqrt{\mathcal J_-\mathcal
J_+}-\sqrt{J_-  J_+})( \mathcal M_4^-)^n], \end{eqnarray}
\begin{eqnarray}
(H^{2n+1})_{41}&=&(\delta/2)J_-\mathcal J_-\frac{(\mathcal
M_1^+)^n-(\mathcal M_1^-)^n}{\sqrt{J_+J_-\mathcal
 J_+\mathcal
J_-}}, \\
(H^{2n+1})_{42}&=&J_-\frac{\alpha/2}{\sqrt{N J_+
J_-}}[(\sqrt{\mathcal J_-\mathcal J_+}+\sqrt{ J_+
 J_-}) (\mathcal M_2^+)^n  \\&+&(\sqrt{J_+  J_-}-\sqrt{\mathcal J_-\mathcal
J_+})(\mathcal M_2^-)^{n}], \\
(H^{2n+1})_{43}&=&\mathcal  J_-\frac{\alpha/2}{\sqrt{N \mathcal
J_+\mathcal J_-}}[(\sqrt{\mathcal J_+\mathcal J_-}+\sqrt{J_-
J_+})(\mathcal M_3^+)^n \\&+&(\sqrt{\mathcal J_+\mathcal
J_-}-\sqrt{J_-  J_+})( \mathcal M_3^-)^n],  \\
(H^{2n+1})_{44}&=&\frac{1}{2}\delta[(\mathcal M_4^+)^n+(\mathcal
M_4^-)^n].\label{powerodd2}
\end{eqnarray}
Having in hand the explicit expressions of  powers of the total
Hamiltonian, it can easily be verified that the elements of the time
evolution operator, obtained by inserting equations~(\ref{power1})
and~(\ref{powerodd1})-(\ref{powerodd2}) into
equation~(\ref{taylor}), are given by
\begin{align}
U_{11}(t)=&\frac{1}{2}\Bigl\{\cos\Bigl(t \sqrt{ \mathcal
M_1^+}\Bigl)+\cos\Bigl(t\sqrt{ \mathcal
M_1^-}\Bigl)-i\delta\Bigl[\frac{\sin\Bigl(t\sqrt{ \mathcal
M_1^+}\Bigl)}{\sqrt{ \mathcal M_1^+}}+\frac{\sin\Bigl(t \sqrt{
\mathcal M_1^-}\Bigl)}{ \sqrt{ \mathcal M_1^-}}\Bigl]\Bigl\},\\
U_{21}(t)=&-\mathcal J_-\frac{i\alpha/2}{\sqrt{N\mathcal J_+\mathcal
J_-}}\Bigl\{\frac{\sqrt{J_+J_-}+\sqrt{\mathcal J_+ \mathcal
J_-}}{\sqrt{ \mathcal M_1^+}}\sin\Bigl(t\sqrt{ \mathcal
M_1^+}\Bigl)\nonumber\\&-\frac{\sqrt{J_+J_-}-\sqrt{\mathcal J_+
\mathcal J_-}}{\sqrt{ \mathcal M_1^-}}\sin\Bigl(t \sqrt{ \mathcal
M_1^-}\Bigl)\Bigl\},\\
U_{31}(t)=&- J_-\frac{i\alpha/2}{\sqrt{N
J_+J_-}}\Bigl\{\frac{\sqrt{J_+J_-}+\sqrt{\mathcal J_+ \mathcal
J_-}}{\sqrt{ \mathcal M_1^+}}\sin\Bigl(t\sqrt{ \mathcal
M_1^+}\Bigl)\nonumber\\&+\frac{\sqrt{J_+J_-}-\sqrt{\mathcal J_+
\mathcal J_-}}{\sqrt{ \mathcal M_1^-}}\sin\Bigl(t \sqrt{ \mathcal
M_1^-}\Bigl)\Bigl\},\end{align} \begin{align} U_{41}(t)=&J_-\mathcal
J_-\frac{1}{2\sqrt{J_+J_-\mathcal J_+\mathcal
J_-}}\Bigl\{\cos\Bigl(t \sqrt{ \mathcal
M_1^+}\Bigl)-\cos\Bigl(t\sqrt{ \mathcal
M_1^-}\Bigl)\nonumber\\&-i\delta\Bigl[\frac{\sin\Bigl(t\sqrt{
\mathcal M_1^+}\Bigl)}{\sqrt{ \mathcal M_1^+}}-\frac{\sin\Bigl(t
\sqrt{ \mathcal M_1^-}\Bigl)}{ \sqrt{ \mathcal
M_1^-}}\Bigl]\Bigl\},\\ U_{22}(t)=&\frac{1}{2}\Bigl\{\cos\Bigl(t
\sqrt{ \mathcal M_2^+}\Bigl)+\cos\Bigl(t\sqrt{ \mathcal
M_1^-}\Bigl)+i\delta\Bigl[\frac{\sin\Bigl(t\sqrt{ \mathcal
M_2^+}\Bigl)}{\sqrt{ \mathcal M_2^+}}+\frac{\sin\Bigl(t \sqrt{
\mathcal M_2^-}\Bigl)}{ \sqrt{ \mathcal M_2^-}}\Bigl]\Bigl\},\\
U_{12}(t)=&-\mathcal J_+\frac{i\alpha/2}{\sqrt{N\mathcal J_-\mathcal
J_+}}\Bigl\{\frac{\sqrt{J_+J_-}+\sqrt{\mathcal J_- \mathcal
J_+}}{\sqrt{ \mathcal M_2^+}}\sin\Bigl(t\sqrt{ \mathcal
M_2^+}\Bigl)\nonumber\\&-\frac{\sqrt{J_+J_-}-\sqrt{\mathcal J_-
\mathcal J_+}}{\sqrt{ \mathcal M_2^-}}\sin\Bigl(t \sqrt{ \mathcal
M_2^-}\Bigl)\Bigl\},\\
U_{32}(t)=&J_-\mathcal J_+\frac{1}{2\sqrt{J_+J_-\mathcal J_-\mathcal
J_+}}\Bigl\{\cos\Bigl(t \sqrt{ \mathcal
M_2^+}\Bigl)-\cos\Bigl(t\sqrt{ \mathcal
M_2^-}\Bigl)\nonumber\\&+i\delta\Bigl[\frac{\sin\Bigl(t\sqrt{
\mathcal M_2^+}\Bigl)}{\sqrt{ \mathcal M_2^+}}-\frac{\sin\Bigl(t
\sqrt{ \mathcal M_2^-}\Bigl)}{ \sqrt{ \mathcal
M_2^-}}\Bigl]\Bigl\},\\
 U_{42}(t)=&- J_-\frac{i\alpha/2}{\sqrt{N
J_+J_-}}\Bigl\{\frac{\sqrt{J_+J_-}+\sqrt{\mathcal J_- \mathcal
J_+}}{\sqrt{ \mathcal M_2^+}}\sin\Bigl(t\sqrt{ \mathcal
M_2^+}\Bigl)\nonumber\\&+\frac{\sqrt{J_+J_-}-\sqrt{\mathcal J_-
\mathcal J_+}}{\sqrt{ \mathcal M_2^-}}\sin\Bigl(t \sqrt{ \mathcal
M_2^-}\Bigl)\Bigl\},\\
 U_{33}(t)=&\frac{1}{2}\Bigl\{\cos\Bigl(t
\sqrt{ \mathcal M_3^+}\Bigl)+\cos\Bigl(t\sqrt{ \mathcal
M_3^-}\Bigl)+i\delta\Bigl[\frac{\sin\Bigl(t\sqrt{ \mathcal
M_3^+}\Bigl)}{\sqrt{ \mathcal M_3^+}}+\frac{\sin\Bigl(t \sqrt{
\mathcal M_3^-}\Bigl)}{ \sqrt{ \mathcal M_3^-}}\Bigl]\Bigl\},\\
U_{13}(t)=&- J_+\frac{i\alpha/2}{\sqrt{N J_-
J_+}}\Bigl\{\frac{\sqrt{\mathcal J_+\mathcal J_-}+\sqrt{ J_-
 J_+}}{\sqrt{ \mathcal M_3^+}}\sin\Bigl(t\sqrt{ \mathcal
M_3^+}\Bigl)\nonumber\\&-\frac{\sqrt{\mathcal J_+\mathcal
J_-}-\sqrt{ J_-  J_+}}{\sqrt{ \mathcal M_3^-}}\sin\Bigl(t \sqrt{
\mathcal
M_3^-}\Bigl)\Bigl\},\\
U_{23}(t)=&J_+\mathcal J_- \frac{1}{2\sqrt{J_- J_+ \mathcal
J_+\mathcal J_- }}\Bigl\{\cos\Bigl(t \sqrt{ \mathcal
M_3^+}\Bigl)-\cos\Bigl(t\sqrt{ \mathcal
M_3^-}\Bigl)\nonumber\\&+i\delta\Bigl[\frac{\sin\Bigl(t\sqrt{
\mathcal M_3^+}\Bigl)}{\sqrt{ \mathcal M_3^+}}-\frac{\sin\Bigl(t
\sqrt{ \mathcal M_3^-}\Bigl)}{ \sqrt{ \mathcal
M_3^-}}\Bigl]\Bigl\},\\
 U_{43}(t)=&- \mathcal J_-\frac{i\alpha/2}{\sqrt{N
\mathcal J_+\mathcal J_-}}\Bigl\{\frac{\sqrt{\mathcal J_+\mathcal
J_-}+\sqrt{ J_-  J_+}}{\sqrt{ \mathcal M_3^+}}\sin\Bigl(t\sqrt{
\mathcal M_3^+}\Bigl)\nonumber\\&+\frac{\sqrt{\mathcal J_+\mathcal
J_-}-\sqrt{ J_- J_+}}{\sqrt{ \mathcal M_3^-}}\sin\Bigl(t \sqrt{
\mathcal M_3^-}\Bigl)\Bigl\},\end{align}
\begin{align}
U_{44}(t)=&\frac{1}{2}\Bigl\{\cos\Bigl(t \sqrt{ \mathcal
M_4^+}\Bigl)+\cos\Bigl(t\sqrt{ \mathcal
M_4^-}\Bigl)-i\delta\Bigl[\frac{\sin\Bigl(t\sqrt{ \mathcal
M_4^+}\Bigl)}{\sqrt{ \mathcal M_4^+}}+\frac{\sin\Bigl(t \sqrt{
\mathcal M_4^-}\Bigl)}{ \sqrt{ \mathcal M_4^-}}\Bigl]\Bigl\},\\
U_{24}(t)=&- J_+\frac{i\alpha/2}{\sqrt{N J_-
J_+}}\Bigl\{\frac{\sqrt{\mathcal J_-\mathcal J_+}+\sqrt{ J_-
 J_+}}{\sqrt{ \mathcal M_4^+}}\sin\Bigl(t\sqrt{ \mathcal
M_4^+}\Bigl)\nonumber\\&-\frac{\sqrt{\mathcal J_-\mathcal
J_+}-\sqrt{J_-  J_+}}{\sqrt{ \mathcal M_4^-}}\sin\Bigl(t \sqrt{
\mathcal
M_4^-}\Bigl)\Bigl\},\\
U_{34}(t)=&-\mathcal  J_+\frac{i\alpha/2}{\sqrt{N \mathcal
J_-\mathcal J_+}}\Bigl\{\frac{\sqrt{\mathcal J_-\mathcal
J_+}+\sqrt{J_-  J_+}}{\sqrt{ \mathcal M_4^+}}\sin\Bigl(t\sqrt{
\mathcal M_4^+}\Bigl)\nonumber\\&+\frac{\sqrt{\mathcal J_-\mathcal
J_+}-\sqrt{J_-  J_+}}{\sqrt{ \mathcal M_4^-}}\sin\Bigl(t \sqrt{
\mathcal
M_4^-}\Bigl)\Bigl\},\\
U_{14}(t)=&J_+\mathcal J_+\frac{1}{2\sqrt{J_- J_+\mathcal
J_-\mathcal J_+}}\Bigl\{\cos\Bigl(t \sqrt{ \mathcal
M_4^+}\Bigl)-\cos\Bigl(t\sqrt{ \mathcal
M_4^-}\Bigl)\nonumber\\&-i\delta\Bigl[\frac{\sin\Bigl(t\sqrt{
\mathcal M_4^+}\Bigl)}{\sqrt{ \mathcal M_4^+}}-\frac{\sin\Bigl(t
\sqrt{ \mathcal M_4^-}\Bigl)}{ \sqrt{ \mathcal M_4^-}}\Bigl]\Bigl\}.
\end{align}

 It should be noted that the above components of the operator
$\mathbf U(t)$ can also be derived by solving the Schr\"{o}dinger
equation~\cite{22}
\begin{equation}
i\frac{d \mathbf U(t)}{dt}=H{\mathbf U}(t).\end{equation} For
instance, we have
\begin{align}
i \frac{d U_{11}(t)}{dt}=\delta
U_{11}(t)+\frac{\alpha}{\sqrt{N}}\mathcal J_+ U_{21}(t)+
\frac{\alpha}{\sqrt{N}}J_+U_{31}(t)\label{case2eq1},\\
i \frac{d U_{21}(t)}{dt}=\frac{\alpha}{\sqrt{N}}\mathcal J_-
U_{11}(t)-\delta U_{21}(t)+
\frac{\alpha}{\sqrt{N}}J_+U_{41}(t),\\
i \frac{d U_{31}(t)}{dt}=\frac{\alpha}{\sqrt{N}} J_-U_{11}(t)-\delta
U_{31}(t)+
\frac{\alpha}{\sqrt{N}}\mathcal J_+ U_{41}(t),\\
i \frac{d U_{41}(t)}{dt}=\frac{\alpha}{\sqrt{N}} J_- U_{21}(t)+
\frac{\alpha}{\sqrt{N}}\mathcal J_-U_{31}(t)+\delta
U_{41}(t)\label{case2eq4}.\end{align} This set of differential
equation can be solved by introducing the following transformations:
 \begin{align}
 U_{11}(t)&\rightarrow e^{-i\delta t}U_{11}(t),\\
U_{21}(t)&\rightarrow e^{-i\delta t}\mathcal J_- U_{21}(t),\\
U_{31}(t)&\rightarrow e^{-i\delta t}J_-U_{31}(t), \\
U_{41}(t)&\rightarrow e^{-i\delta t}J_-\otimes \mathcal
J_-U_{41}(t).
\end{align}
The resulting  differential equations involve diagonal terms; they
can be  solved by taking into account the initial conditions:
\begin{equation}
U_{ij}(0)=\left\{
            \begin{array}{ll}
              {\mathbf 1}    &\text{for} \quad \hbox{$i=j$,} \\
              0 &  \text{for} \quad\hbox{$i\neq j$.}
            \end{array}
          \right.
\end{equation}
 Following the same procedure, it is possible to derive the remaining matrix elements of the time evolution operator.

 There exist many measures for entanglement. Here we shall use the concurrence, defined
 by~\cite{26}
\begin{equation}
C(\rho)=\max\{0,2\max[\sqrt{\lambda_i}]-\sum_{i=1}^4\sqrt{\lambda_i}\},\end{equation}
where the quantities  $\lambda_i$ are the eigenvalues of the
operator
$\rho(t)(\sigma_y\otimes\sigma_y)\rho(t)^*(\sigma_y\otimes\sigma_y)$.
The above measure is equal to one for maximally entangled states,
and is equal to zero for separable states. The purity
\begin{equation} P(t)={\rm tr}\rho(t)^2\end{equation} can be used
to quantify the decoherence of the central system; it is equal to
$\tfrac{1}{4}$ for maximally mixed states, and one for pure states.

 It turns out that the density matrices corresponding to the
initial product states $|\epsilon_1\epsilon_2\rangle$, where
$\epsilon_i\equiv\pm$, are always diagonal. Furthermore, the
numerical simulation shows that if the qubits are prepared in one of
the above states, they remain unentangled regardless of the values
of $N$ and $\delta$. The purity decays less with the increase of
$\delta$.
 \begin{figure}[htba]
{\centering
\resizebox*{0.65\textwidth}{!}{\includegraphics{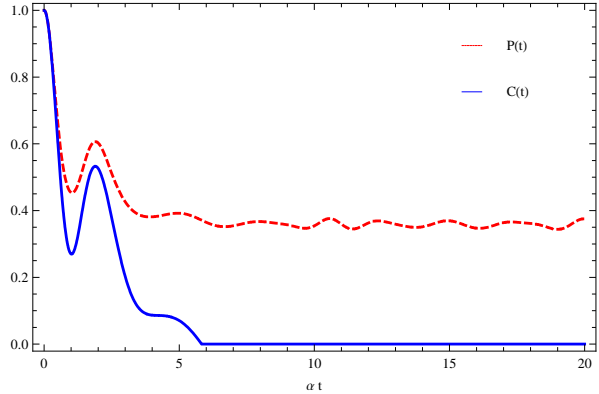}}
\par}

\caption{\label{figure1}  The evolution in time of the concurrence
(solid curve) and the purity (dashed curve) corresponding to the
singlet state for $\delta=\alpha$ and $N=10$.}
\end{figure}
\begin{figure}[htba]
{\centering
\resizebox*{0.65\textwidth}{!}{\includegraphics{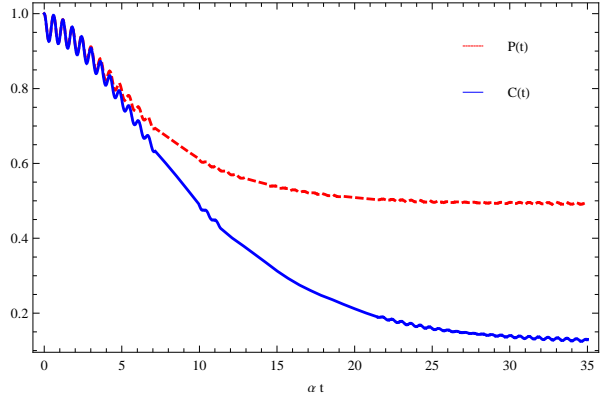}}
\par}

\caption{\label{figure2}  The evolution in time of the concurrence
(solid curve) and the purity (dashed curve) corresponding to the
singlet state for $\delta=4\alpha$ and $N=10$.}
\end{figure}

The matrix elements of the reduced density matrices corresponding to
the states $\tfrac{1}{\sqrt{2}}(|-+\rangle\pm|+-\rangle)$ and
$\tfrac{1}{\sqrt{2}}(|++\rangle\pm|--\rangle)$ are shown in the
Appendix.  The evolution in time of the concurrence and the purity
corresponding to the above maximally entangled  states is
practically the same; we only present the results obtained for the
singlet state. It is found that, for fixed $\delta$, the concurrence
and the purity saturate as the number of spins increases. This
naturally suggests the investigation of the case $N\to\infty$ (see
the next section). For small values of the coupling constant
$\delta$, the concurrence decays from its initial maximum value
$C_{\mathrm{max}}=1$, then vanishes at a certain moment of time
(i.e. entanglement sudden death~\cite{27}).  At long times, and
sufficiently large $N$ and $\delta$, the purity and the concurrence
converge to certain asymptotic values, which increase with the
increase of the strength of interaction. Here it should be noted
that, in contrast to
 the case of common spin bath, the
singlet state is not decoherence free. This was expected because the
latter state  is not eigenvector of the Hamiltonian $H$.
Nevertheless, we find that decoherence can be reduced
 with  strong coupling between the qubits, in agreement with~\cite{22}. Finally let us remark that,
  although we only have considered infinite temperature, we can
 ensure that  for long-range antiferromagnetic Heisenberg interactions within the baths,
    low  temperatures  will have the same effect
on decoherence and entanglement of the qubits as strong coupling
constants.

\section{Thermodynamic limit\label{sec4}}
In the thermodynamic limit, $N\to\infty$, the operators $\sqrt{J_\pm
J_\mp/N}$ converge to the positive real random variable $r$ whose
 probability density function is given by
\begin{equation}
r\mapsto f(r)=4 r {\rm e}^{-2 r^2},  \quad  r \ge
0.\label{prod5}\end{equation} Indeed, it has been shown
in~\cite{22,23} that the operator $J_+/\sqrt{N}$ converges to the
complex  normal random variable $z$ with the  probability density
function
\begin{equation}
z\mapsto \frac{2}{\pi} \mathrm e^{-2|z|^2}.
\end{equation}
Expressing
 $z$ in terms of the polar coordinates $r$ and $\phi$, i.e.,
$z=r \mathrm e^{i \phi}$,  simply gives $|z|^2=r^2$. Then
integrating the corresponding probability density function over the
variable $\phi$ from 0 to $2 \pi$ yields
\begin{eqnarray}
 dP(r)=f(r)dr&=&\frac{2}{\pi} \int\limits_0^{2\pi} d\phi \ r \ dr \mathrm e^{-2
 r^2}\nonumber\\
 &=&4 r \mathrm e^{-2
 r^2} \ dr,
\end{eqnarray}
from which~(\ref{prod5}) follows.

Hence we can ascertain that
\begin{equation}
\lim\limits_{N\to\infty} 2^{-2N}{\rm tr}_{B_1+B_2}
\Omega\Bigl(\sqrt{J_\pm J_\mp/N},\sqrt{\mathcal J_\pm\mathcal
J_\mp/N}\Bigl)=16\int\limits^{\infty}_{0}\int\limits^{\infty}_{0} r\
s\ \mathrm e^{-2(r^2+s^2)}\Omega(r,s) dr
ds,\label{sumint}\end{equation} where $\Omega(r,s)$ is some
complex-valued function for which the integrals in the right-hand
side of equation~(\ref{sumint}) converge.

 Using the above result, one can  express the nonzero elements of the reduced density matrix
 corresponding  to the initial state
$\frac{1}{\sqrt{2}}(|-+\rangle-|+-\rangle)$, in the thermodynamic
limit, as
\begin{eqnarray}
\rho_{11}(t)&=&\rho_{44}(t)=\frac{1}{4}[\Lambda_+(t)+\Lambda_-(t)],\\
\rho_{22}(t)&=&\rho_{33}(t)=\frac{1}{4}[\Upsilon_+(t)+\Upsilon_-(t)+\Xi_+(t)+\Xi_-(t)],\\
\rho_{23}(t)&=&-\frac{1}{8}[\Upsilon_+(t)+\Upsilon_-(t)+\Xi_+(t)+\Xi_-(t)+2\Psi(t)],
\end{eqnarray}
where ( we set $\alpha=1$ for the sake of shortness)
\begin{eqnarray}
\Lambda_\pm(t)&=&16\int\limits_0^\infty\int\limits_0^\infty  r s \
\mathrm e^{-2(r^2+s^2)} \frac{(r\pm s)^2}{\delta^2+(r\pm
s)^2}\sin^2\Bigl(t \sqrt{\delta^2+(r\pm s)^2}\Bigl)dr ds,\label{f1}\\
\Upsilon_\pm(t)&=&16\int\limits_0^\infty\int\limits_0^\infty  r s \
\mathrm e^{-2(r^2+s^2)} \cos^2\Bigl(t \sqrt{\delta^2+(r\pm
s)^2}\Bigl)dr ds,\label{f2}\\
\Xi_\pm(t)&=& 16\int\limits_0^\infty\int\limits_0^\infty  r s \
\mathrm e^{-2(r^2+s^2)} \frac{\delta^2}{\delta^2+(r\pm
s)^2}\sin^2\Bigl(t
\sqrt{\delta^2+(r\pm s)^2}\Bigl)dr ds,\label{f3}\\
 \Psi(t)&=&16\int\limits_0^\infty\int\limits_0^\infty  r s \
\mathrm e^{-2(r^2+s^2)} \Bigl\{\cos\Bigl(t \sqrt{\delta^2+(r+
s)^2}\Bigl) \cos\Bigl(t \sqrt{\delta^2+(r- s)^2}\Bigl)\nonumber
\\&+&\delta^2 \frac{\sin\Bigl(t
\sqrt{\delta^2+(r+ s)^2}\Bigl)}{\delta^2+(r+ s)^2}\frac{\sin\Bigl(t
\sqrt{\delta^2+(r- s)^2}\Bigl)}{\delta^2+(r- s)^2}\Bigl\} dr
ds\label{f4}.
\end{eqnarray}

Unfortunately the above functions cannot be evaluated analytically;
one should make recourse to numerical integration. This task can be
significantly simplified by transforming the double integration into
single one, which is much easier to carry out. To do that notice
that the analysis of the expressions of the functions
$\Lambda_\pm(t)$, $\Upsilon_\pm(t)$, and $\Xi_\pm(t)$  leads  to the
evaluation of the probability density functions $Q(\mu)$ and
$R(\eta)$ corresponding, respectively, to the random variables
$\mu=r+s$ and $\eta=r-s$ (see~\cite{28} for a similar situation).

 Let us begin with the variable $\mu$; its probability density
 function is simply given by the convolution of $f(r)$ with itself:
 \begin{equation}
Q(\mu)=16 \int\limits_{0}^\mu (\mu-r) r \mathrm e^{-2(\mu-r)^2-2
r^2} dr .\end{equation} Note that the upper limit of the integration
over $r$ is $\mu$ because the quantity $\mu-r$ should be positive.
The evaluation of the integral is somewhat lengthy, but elementary;
one finds that
\begin{equation}
Q(\mu)=[2 \mu-\sqrt{\pi}\mathrm e^{\mu^2}(1-2\mu^2) {\rm
erf}(\mu)]\mathrm e^{-2\mu^2},\label{probfun1}\end{equation} where
${\rm erf}(x)$ designates the error function~\cite{29}.

\begin{figure}[htba]
{\centering
\resizebox*{0.60\textwidth}{!}{\includegraphics{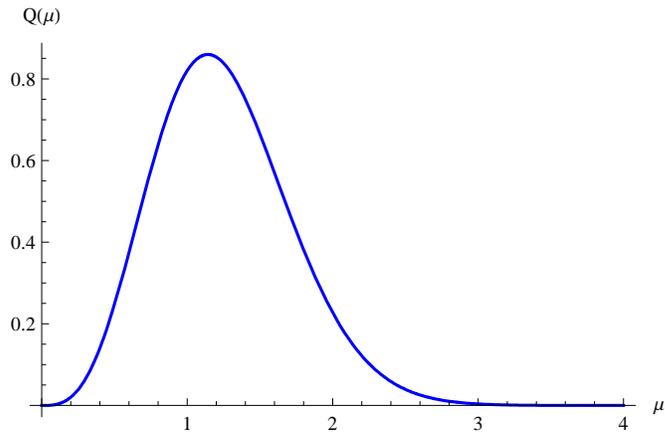}}
\par}

\caption{\label{figure3}  The  probability density function
$Q(\mu)$.}
\end{figure}

\begin{figure}[htba]
{\centering
\resizebox*{0.60\textwidth}{!}{\includegraphics{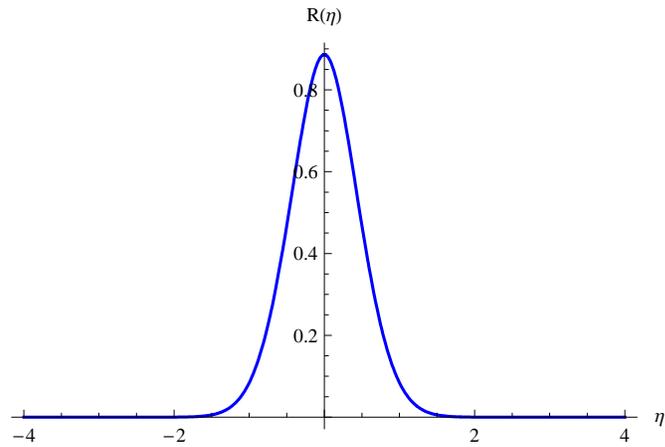}}
\par}

\caption{\label{figure4} The  probability density function
$R(\eta)$.}
\end{figure}

Now consider the variable  $\eta=r-s$. One should be careful when
using the definition of the convolution,
since, in this case,  $\eta$ belongs  to the interval
$]-\infty,\infty[ $. We have to distinguish between two cases,
namely, $\eta\ge 0$  and $\eta\le 0$. In the first case $r\in
[0,\infty[$, and hence
\begin{eqnarray}
R(\eta\ge0)&=&16 \int\limits_0^\infty (\eta+r) r \mathrm
e^{-2(r+s)^2-2r^2} dr\nonumber \\
&=&\frac{1}{2} \{2\eta +\sqrt{\pi} \mathrm e^{\eta^2}  (1 - 2
\eta^2) [1-{\rm erf}(\eta)]\}\mathrm e^{-2 \eta^2}.\label{pp1}
\end{eqnarray}
When $\eta\le 0$, then $r\in[-\eta,\infty[$, which implies that
\begin{eqnarray}
R(\eta\le 0)&=&16 \int\limits_{-\eta}^\infty (\eta+r) r \mathrm
e^{-2(r+s)^2-2r^2} dr\nonumber \\
&=&\frac{1}{2} \{-2\eta +\sqrt{\pi} \mathrm e^{\eta^2}  (1 - 2
\eta^2) [1+{\rm erf}(\eta)]\}\mathrm e^{-2 \eta^2}.\label{pp2}
\end{eqnarray}
Combining (\ref{pp1}) and (\ref{pp2}), we obtain the following
expression for the probability density function of $\eta$ over the
real line:
\begin{eqnarray}
R(\eta)=\frac{1}{2} \{2|\eta| +\sqrt{\pi} \mathrm e^{\eta^2}  (1 - 2
\eta^2) [1-{\rm erf}(|\eta|)]\}\mathrm e^{-2 \eta^2}\label{profun2}.
\end{eqnarray}
The above functions are depicted in figures~\ref{figure3}
and~\ref{figure4}. Clearly, $R(\eta)$ is an even function of its
argument; it takes its maximum value at the origin, that is,
$\max\{R(\eta)\}=R(0)=0.886227$. The maximum value of $Q(\mu)$
occurs at $\mu_0=1.14209$, such that
$\max\{Q(\mu)\}=Q(\mu_0)=0.859664$.

\vline

As a simple application let us prove the following:
\begin{theorem}
 The moments around origin of
the random variables $\mu$ and $\eta$  are given by:
\begin{align}
\langle \mu^{2n} \rangle&=\frac{n!}{2^n}\Bigl[1+2^{n+1} n\
_2F_1\Bigl( 1 + n,\frac{1}{2}; \frac{3}{2}; -1\Bigl)\Bigl],\label{mom1}\\
\langle \mu^{2n+1}
\rangle&=\frac{\Gamma(\tfrac{3}{2}+1)}{2^n}\Bigl[\tfrac{1}{\sqrt{2}}+2^{n}
(2n+1)\ _2F_1\Bigl( \frac{3}{2} + n,\frac{1}{2}; \frac{3}{2};
-1\Bigl)\Bigl],\\
 \langle \eta^{2n}
\rangle&=\langle \mu^{2n}
\rangle-n\sqrt{\pi}\Gamma\Bigl(\frac{1}{2}+n\Bigl),\\
\langle \eta^{2n+1} \rangle&=0,\label{mom4}
\end{align}
where $\Gamma(x)$, and $_2F_1(a,b;c;d)$ denote the Gamma  and the
hypergeometric functions, respectively.
\end{theorem}

\begin{proof}  Relation~(\ref{mom4})
is obvious since the function $R(\eta)$ is even. Let us
prove~(\ref{mom1}). We have that
\begin{eqnarray}
 \langle \mu^{2n} \rangle&=&\int\limits_0^\infty \mu^{2n}Q(\mu)\
 d\mu\nonumber\\
 &=&2I_{n+1}-I_n+2 Y_n,\end{eqnarray}
 where
\begin{eqnarray}
I_n&=&\int\limits_0^\infty\sqrt{\pi} \mu^{2n}  \mathrm e^{-
\mu^2}{\rm erf}(\mu)\ d\mu,\label{funi}\\
Y_n&=&\int\limits_0^\infty \mu^{2n+1} \mathrm e^{-2 \mu^2}\ d\mu.
 \end{eqnarray}
To calculate $Y_n$ and $I_{n}$,  introduce the functions of the real
variable $x>0$:
\begin{equation}
Y_n(x)=\int\limits_0^\infty\mu^{2n+1}\mathrm
e^{-\mu^2(1+\tfrac{1}{x})} d\mu,
\end{equation}
\begin{equation}
I_n(x)=\int\limits_0^\infty\sqrt{\pi} \mu^{2n}  \mathrm e^{-
\mu^2/x}{\rm erf}(\mu)\ d\mu. \label{newfun}\end{equation} The first
integral can be easily evaluated:
 \begin{eqnarray}
\nonumber\\
Y_n(x)&=&\frac{1}{2}\Bigl(\tfrac{x}{1+x}\Bigl)^{n+1}
\int\limits_0^\infty\chi^{n}\mathrm e^{-\chi}d\chi=
\frac{n!}{2}\Bigl(\tfrac{x}{1+x}\Bigl)^{n+1}\label{firsteq}.\end{eqnarray}
The second integral satisfies
\begin{equation}
\frac{d I_n(x)}{dx}=\frac{1}{x^2}
I_{n+1}(x).\label{difffn}\end{equation} Integrating by parts the RHS
of (\ref{newfun})
 with respect to $\mu$, and using~(\ref{firsteq}), yield
\begin{equation}
I_{n+1}(x)=\frac{x(2n+1)}{2}I_n(x)+\frac{x
n!}{2}\Bigl(\frac{x}{x+1}\Bigl)^{n+1}.\label{redn}\end{equation}
Here we have used the fact that ${\rm erf}(x)^{'} =2 \mathrm
e^{-x^2}/\sqrt{\pi}$.

 Let $I_n(x)=n! x^{n+1}  g_n(x)$. Then
from~(\ref{redn}) we have
\begin{equation}
2(n+1)
g_{n+1}(x)=(2n+1)g_n(x)+\frac{1}{(x+1)^{n+1}}\label{dign}.\end{equation}
On the other hand equation~(\ref{difffn}) implies that
\begin{equation}
x \frac{d g_n(x)}{dx}+(n+1)g_n(x)=(n+1)g_{n+1}(x).\end{equation}
Combining the last two equations yields the following first order
differential  equation for the function $g_n(x)$:
\begin{equation}
2 x \frac{d
g_n(x)}{dx}+g_n(x)-\frac{1}{(x+1)^{n+1}}=0\label{difff}.\end{equation}
Differentiating both sides of~(\ref{difff}), and again
using~(\ref{dign}), we obtain \begin{equation}
\Bigl[\frac{d^2}{dx^2}+\Bigl(\frac{3}{2x}+\frac{n+1}{x+1}\Big)\frac{d}{dx}+\frac{n+1}{2x(x+1)}\Bigl]g_n(x)=0.\end{equation}
By setting $y=-x$, and $h_n(y)=g_n(-x)$, we obtain \begin{equation}
\Bigl[\frac{d^2}{dy^2}+\Bigl(\frac{3}{2y}+\frac{n+1}{y-1}\Big)\frac{d}{dy}+\frac{n+1}{2y(y-1)}\Bigl]h_n(y)=0,
\end{equation}
which should be compared with the hypergeometric equation
\begin{equation}
\Bigl[\frac{d^2}{dy^2}+\Bigl(\frac{c}{y}+\frac{1+a+b-c}{y-1}\Big)\frac{d}{dy}+\frac{ab}{y(y-1)}\Bigl]\
_2F_1(a,b;c;y)=0.
\end{equation}
Thus

\begin{equation*}a=n+1,\quad b=\tfrac{1}{2}, \quad
c=\tfrac{3}{2}.\end{equation*} It follows  that
\begin{equation}
I_n(x)=n! x^{n+1} \
_2F_1(n+1,\tfrac{1}{2};\tfrac{3}{2};-x).\end{equation} Putting $x=1$
yields
\begin{equation}
I_n=n!  \ _2F_1(n+1,\tfrac{1}{2};\tfrac{3}{2};-1), \qquad
Y_n=\frac{n!}{2^{n+2}}.\end{equation}
 Also, using~(\ref{redn}), we obtain
\begin{equation}2I_{n+1}=(2n+1)n!  \
_2F_1(n+1,\tfrac{1}{2};\tfrac{3}{2};-1)+\frac{n!}{2^{n+1}},\end{equation}
from which (\ref{mom1}) readily  follows. The other moments  can be
evaluated with a similar method. \hfill
\medskip
\end{proof}

 The functions~(\ref{f1})-(\ref{f3}) can
easily be expressed in terms of the functions $Q(\mu)$ and
$R(\eta)$. For example, we have:
\begin{eqnarray}
\Lambda_+(t)&=&\int_0^\infty
Q(\mu)\frac{\mu^2}{\delta^2+\mu^2}\sin^2\Bigl(t\sqrt{\delta^2+\mu^2}\Bigl)
d\mu,\\
\Lambda_-(t)&=&\int_{-\infty}^\infty
R(\mu)\frac{\mu^2}{\delta^2+\mu^2}\sin^2\Bigl(t\sqrt{\delta^2+\mu^2}\Bigl)
d\mu.
\end{eqnarray}
It should be noted that in contrast to $r$ and $s$, the random
variables $\eta$ and $\mu$ are not independent. The function
$\Psi(t)$ can not be further simplified, and should be evaluated
using the double integration over the variables $r$ and $s$.
Nevertheless, using the Riemann-Lebesgue lemma, we can infer that
\begin{equation}
\lim\limits_{t\to\infty}\Psi(t)=\Psi(\infty)=0.\end{equation} In a
similar way, the remaining functions tend asymptotically to:
\begin{eqnarray}
\Lambda_+(\infty)&=&\frac{1}{2}\int_0^\infty
Q(\mu)\frac{\mu^2}{\delta^2+\mu^2}
d\mu,\\
\Lambda_-(\infty)&=&\frac{1}{2}\int_{-\infty}^\infty
R(\mu)\frac{\mu^2}{\delta^2+\mu^2} d\mu,\\
\Upsilon_\pm(\infty)&=&\frac{1}{2},\\
 \Xi_+(\infty)&=&\frac{1}{2}\int_0^\infty
Q(\mu)\frac{\delta^2}{\delta^2+\mu^2} d\mu,\\
\Xi_-(\infty)&=&\frac{1}{2}\int_{-\infty}^\infty
R(\mu)\frac{\delta^2}{\delta^2+\mu^2} d\mu.
\end{eqnarray}
Notice that \begin{equation}
\Lambda_\pm(\infty)+\Xi_\pm(\infty)=\frac{1}{2},
\end{equation}
independently of the values of $\delta$. It follows that the
asymptotic density matrix can be expressed as
\begin{equation}
\rho(\infty)=\begin{pmatrix} \frac{\Pi}{4}&&0&&0&&0\\
0&& \frac{2-\Pi}{4}&&-\frac{2-\Pi}{8}&&0\\
0&&-\frac{2-\Pi}{8}&&\frac{2-\Pi}{4}&&0\\0&&0&&0&&\frac{\Pi}{4}\end{pmatrix},\end{equation}
where
\begin{equation}
\Pi=\Lambda_+(\infty)+\Lambda_-(\infty).\end{equation} It is easily
seen that
\begin{equation}
\lim\limits_{\delta\to 0}\Xi_\pm(\infty)=0, \quad
\lim\limits_{\delta\to 0}\Lambda_\pm(\infty)=\frac{1}{2}.
\end{equation}
The corresponding asymptotic reduced density matrix reads
\begin{equation}
\rho(\infty)_{\delta=0}=\begin{pmatrix} \frac{1}{4}&&0&&0&&0\\
0&& \frac{1}{4}&&-\frac{1}{8}&&0\\
0&&-\frac{1}{8}&&\frac{1}{4}&&0\\0&&0&&0&&\frac{1}{4}\end{pmatrix},\end{equation}
which has a concurrence  identically equal to zero.

 On the contrary, in the limit of strong coupling between the central
 qubits,
 \begin{equation}
\lim\limits_{\delta\to \infty}\Xi_\pm(\infty)=\frac{1}{2}, \quad
\lim\limits_{\delta\to \infty}\Lambda_\pm(\infty)=0.\end{equation}
Consequently,
\begin{equation}
\rho(\infty)_{\delta=\infty}=\begin{pmatrix} 0&&0&&0&&0\\
0&& \frac{1}{2}&&-\frac{1}{4}&&0\\
0&&-\frac{1}{4}&&\frac{1}{2}&&0\\0&&0&&0&&0\end{pmatrix}.\end{equation}
A straightforward calculation shows that
\begin{equation}
\lim\limits_{\delta\to\infty}C(\rho(\infty))=\frac{1}{2}.\end{equation}

In general, since $0\le \mu^2/(\mu^2+\delta^2)\le 1$, then
\begin{eqnarray}
0\le \Pi&=\frac{1}{2} \int_0^\infty
Q(\mu)\frac{\mu^2}{\delta^2+\mu^2} d\mu+ \frac{1}{2}
\int_{-\infty}^\infty R(\mu)\frac{\mu^2}{\delta^2+\mu^2} d\mu
\nonumber
\\
&\le \frac{1}{2}\int_0^\infty Q(\mu) d\mu+\frac{1}{2}
\int_{-\infty}^\infty R(\mu) d\mu=1.
\end{eqnarray}
This allows us to find the following explicit form of the asymptotic
value of the concurrence:
\begin{equation}
C(\infty)=\max\Bigl\{0,\frac{2-3\Pi}{4}\Bigl\}.
\end{equation}
The latter can also be rewritten as:
\begin{equation}
C(\infty)=\left\{
            \begin{array}{ll}
              \frac{2-3\Pi}{4} &\text{for} \quad \hbox{$0 \le \Pi \le \frac{2}{3} $,} \\
              0 &  \text{for} \quad\hbox{$ \frac{2}{3}\le \Pi \le 1 $.}
            \end{array}
          \right.
\end{equation}

 The variation of the asymptotic
concurrence as a function of $\delta$ is shown in
figure~\ref{figure5}. It can be seen that $C(\infty)$ remains zero
up to a critical value $\delta_c$ after which it increases, to tend
asymptotically to $\frac{1}{2}$. The value of $\delta_c$ can be
evaluated numerically:
 \begin{figure}[htba]
{\centering
\resizebox*{0.65\textwidth}{!}{\includegraphics{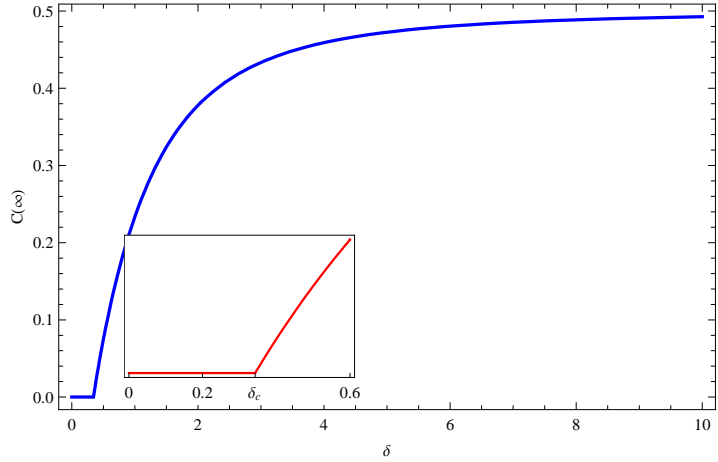}}
\par}

\caption{\label{figure5}  The variation  of $C(\infty)$ as a
function of the coupling constant $\delta$. The inset shows the
critical point $\delta_c$.}
\end{figure}

 \begin{equation}\delta_c= 0.342842, \qquad \Pi|_{\delta=\delta_c}=0.666667.\end{equation}
 At the critical point, the density
matrix reads
\begin{equation}
\rho_c(\infty)=\begin{pmatrix} \frac{1}{6}&&0&&0&&0\\
0&& \frac{1}{3}&&-\frac{1}{6}&&0\\
0&&-\frac{1}{6}&&\frac{1}{3}&&0\\0&&0&&0&&\frac{1}{6}\end{pmatrix}.\end{equation}

\section{Second-order master equation\label{sec5}}
Under Born Approximation, the second-order master equation yields
the following set of integro-differential equations:
\begin{align}
\dot{\tilde{\rho}}_{11}(t)&=-\alpha^2\int\limits_0^t\Big(2\tilde{\rho}_{11}(s)-\tilde{\rho}_{22}(s)
-\tilde{\rho}_{33}(s)\Bigr)\cos[2\delta(t-s)]\ ds,\label{master1}\\
\dot{\tilde{\rho}}_{12}(t)&=-\alpha^2\int\limits_0^t\Big(2\tilde{\rho}_{12}(s)\mathrm
e^{2i\delta(t-s)}-\tilde{\rho}_{34}(s) \mathrm e^{2i\delta(t+s)}\Bigr)\ ds,\label{master2}\\
\dot{\tilde{\rho}}_{13}(t)&=-\alpha^2\int\limits_0^t\Big(2\tilde{\rho}_{13}(s)\mathrm
e^{2i\delta(t-s)}-\tilde{\rho}_{24}(s) \mathrm e^{2i\delta(t+s)}\Bigr)\ ds,\label{master3}\\
\dot{\tilde{\rho}}_{14}(t)&=-\alpha^2\int\limits_0^t
2\tilde{\rho}_{13}(s) \cos[2\delta(t-s)]\ ds,\label{master4}\\
\dot{\tilde{\rho}}_{22}(t)&=-\alpha^2\int\limits_0^t\Big(2\tilde{\rho}_{22}(s)-\tilde{\rho}_{11}(s)
-\tilde{\rho}_{44}(s)\Bigr)\cos[2\delta(t-s)]\ ds,\label{master5}
\end{align}
\begin{align}
\dot{\tilde{\rho}}_{23}(t)&=-\alpha^2\int\limits_0^t
2\tilde{\rho}_{23}(s) \cos[2\delta(t-s)]\ ds,\label{master6}\\
\dot{\tilde{\rho}}_{24}(t)&=-\alpha^2\int\limits_0^t\Big(2\tilde{\rho}_{24}(s)\mathrm
e^{2i\delta(s-t)}-\tilde{\rho}_{13}(s) \mathrm e^{-2i\delta(t+s)}\Bigr)\ ds,\label{master7}\\
\dot{\tilde{\rho}}_{33}(t)&=-\alpha^2\int\limits_0^t\Big(2\tilde{\rho}_{33}(s)-\tilde{\rho}_{11}(s)
-\tilde{\rho}_{44}(s)\Bigr)\cos[2\delta(t-s)]\ ds,\label{master8}\\
\dot{\tilde{\rho}}_{34}(t)&=-\alpha^2\int\limits_0^t\Big(2\tilde{\rho}_{34}(s)\mathrm
e^{2i\delta(s-t)}-\tilde{\rho}_{12}(s) \mathrm e^{-2i\delta(t+s)}\Bigr)\ ds,\label{master9}\\
\dot{\tilde{\rho}}_{44}(t)&=-\alpha^2\int\limits_0^t\Big(2\tilde{\rho}_{44}(s)-\tilde{\rho}_{22}(s)
-\tilde{\rho}_{33}(s)\Bigr)\cos[2\delta(t-s)]\ ds\label{master10}.
\end{align}

Some of the above equations can be solved under a time-local
approximation for which the matrix elements $\tilde{\rho}_{ij}(s)$
are replaced by $\tilde{\rho}_{ij}(t)$. One can find that ($\delta$
and $t$ given in units of $\alpha^{-1}$ and $\alpha$ respectively)
\begin{align}
\tilde{\rho}_{11}(t)&=\frac{1}{4}\Biggl\{1+\Bigl[-1+2(\rho_{11}^0+\rho_{44}^0)\Bigr]\exp\Bigl\{\frac{1}{\delta^2}[\cos(2
\delta t)-1]\Bigl\}\nonumber \\&\qquad +2(\rho_{11}^0-\rho_{44}^0)\exp\Bigl\{\frac{1}{2\delta^2}[\cos(2 \delta t)-1]\Bigr\}\Biggr\},\label{ms11} \\
\tilde{\rho}_{22}(t)&=\frac{1}{4}\Biggl\{1+\Bigl[-1+2(\rho_{22}^0+\rho_{33}^0)\Bigr]\exp\Bigl\{\frac{1}{\delta^2}[\cos(2
\delta t)-1]\Bigl\}\nonumber \\&\qquad +2(\rho_{22}^0-\rho_{33}^0)\exp\Bigl\{\frac{1}{2\delta^2}[\cos(2 \delta t)-1]\Bigr\}\Biggr\}, \\
\tilde{\rho}_{33}(t)&=\frac{1}{4}\Biggl\{1+\Bigl[-1+2(\rho_{33}^0+\rho_{22}^0)\Bigr]\exp\Bigl\{\frac{1}{\delta^2}[\cos(2
\delta t)-1]\Bigl\}\nonumber \\&\qquad
+2(\rho_{33}^0-\rho_{22}^0)\exp\Bigl\{\frac{1}{2\delta^2}[\cos(2
\delta t)-1]\Bigr\}\Biggr\},
\end{align}
\begin{align}
\tilde{\rho}_{44}(t)&=\frac{1}{4}\Biggl\{1+\Bigl[-1+2(\rho_{44}^0+\rho_{11}^0)\Bigr]\exp\Bigl\{\frac{1}{\delta^2}[\cos(2
\delta t)-1]\Bigl\}\nonumber  \\&\qquad +2(\rho_{44}^0-\rho_{11}^0)\exp\Bigl\{\frac{1}{2\delta^2}[\cos(2 \delta t)-1]\Bigr\}\Biggr\}, \\
\tilde{\rho}_{14}(t)&=\rho_{14}^0\exp\Bigl\{\frac{1}{\delta^2}[\cos(2
\delta t)-1]\Bigr\},\\
\tilde{\rho}_{23}(t)&=\rho_{23}^0\exp\Bigl\{\frac{1}{\delta^2}[\cos(2
\delta t)-1]\Bigr\}\label{ms23}.
\end{align}
These solutions describe approximately the dynamics at short times.
In fact, the smaller the coupling constant $\delta$, the better
these solutions are.
\begin{figure}[htba]
{\centering
\resizebox*{0.65\textwidth}{!}{\includegraphics{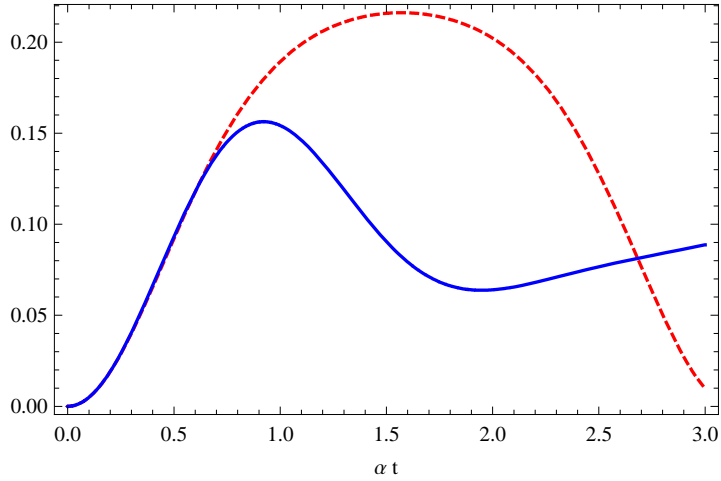}}
\par}

\caption{\label{figure5}  The variation in time  of the  the matrix
element $\rho_{11}(t)$ corresponding to the singlet state. The solid
curve represents the exact solution, and the dashed curve represents
the approximate solution~(\ref{ms11}). The parameters are $N=10$ and
$\delta=\alpha$.}
\end{figure}

\begin{figure}[htba]
{\centering
\resizebox*{0.65\textwidth}{!}{\includegraphics{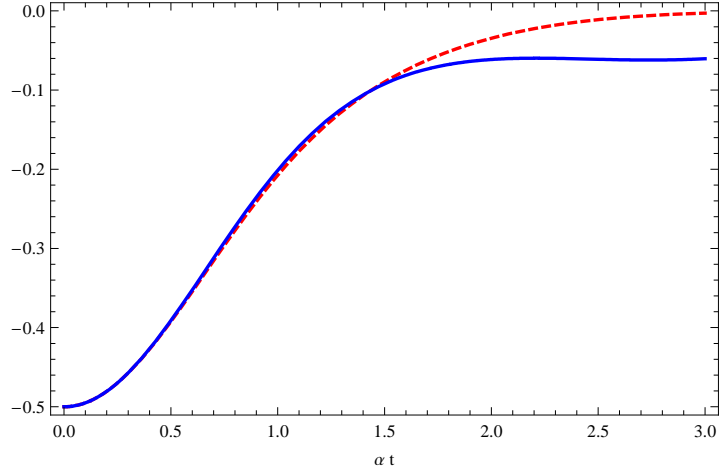}}
\par}

\caption{\label{figure7}  The variation in time  of the  the matrix
element $\rho_{12}(t)$ corresponding to the singlet state. The solid
curve represents the exact solution, and the dashed curve represents
the approximate solution~(\ref{ms12}). The parameters are $N=10$ and
$\delta=0$.}
\end{figure}

Note that when $\delta=0$ ( i.e.  nonlocal dynamics), then
\begin{equation}
\exp\{\frac{1}{n\delta^2}[\cos(2 \delta t)-1]\Bigr\}\to \mathrm
e^{-2 t^2/n}, \qquad n=1,2. \end{equation} Thus the second
 order time-local master equation shows that the nonlocal dynamics,
 or, in general, the short time behavior follow a Gaussian decay law.
Note that the solutions corresponding to the diagonal elements
reproduce  their asymptotic limit, namely,
 $\rho_{ii}(\infty)=\frac{1}{4}$. However, those corresponding to the off-diagonal elements fail to reproduce
 the steady state, since, for
 example, equation~(\ref{ms23}) implies that $\rho_{23}(t)\to 0$. To
 end our discussion let us remark that
equations~(\ref{master2}), (\ref{master3}), (\ref{master7}) and
(\ref{master9}) can be analytically solved  only when $\delta=0$.
For instance (see figure~\ref{figure7}),
\begin{align}
\rho_{12}(t)=\frac{1}{2}\Bigl[(\rho_{12}^0+\rho_{34}^0)\mathrm
e^{-t^2/2}+(\rho_{12}^0-\rho_{34}^0)\mathrm
e^{-3t^2/2}\Bigr].\label{ms12}
\end{align}

\section{Summary}
In summary we have investigated the dynamics of two qubits  coupled
to  separate  spin star environment via Heisenberg $XY$
interactions. We have derived the exact form of the time evolution
operator  and calculated the matrix elements of the reduced density
operator. The analysis of the evolution in time of the concurrence
and the purity shows that decoherence can be minimized by allowing
the central qubits to strongly interact with each other.  The
short-time behavior, studied by deriving the second-order master
equation, is found to be Gaussian.
 The next step may consist in considering more central qubits, and
 investigate whether the above results still hold.
\section*{Appendix}

Using trace properties of the lowering and raising operators, it can
be shown that the nonzero matrix elements corresponding  to the
initial maximally entangled states
$\frac{1}{\sqrt{2}}(|-+\rangle\pm|+-\rangle)$ are explicitly given
by:
\begin{eqnarray}
\rho_{11}(t)&=&2^{-(2N+1)}\mathrm{tr}_{B_1+B_2}\Bigl\{U_{12}(t)U^\dag_{12}(t)+ U_{13}(t)U_{13}^\dag(t)\Bigl\},\\
\rho_{22}(t)&=&2^{-(2N+1)}\mathrm{tr}_{B_1+B_2}\Bigl\{U_{22}(t)U_{22}^\dag(t)+ U_{23}(t)U_{23}^\dag(t)\Bigl\},\\
\rho_{23}(t)&=&\pm 2^{-(2N+1)}\mathrm{tr}_{B_1+B_2}\Bigl\{U_{22}(t)U_{33}^\dag(t)\Bigl\},\\
\rho_{33}(t)&=&2^{-(2N+1)}\mathrm{tr}_{B_1+B_2}\Bigl\{U_{32}(t)U_{32}^\dag(t)+ U_{33}(t)U_{33}^\dag(t)\Bigl\},\\
\rho_{44}(t)&=&2^{-(2N+1)}\mathrm{tr}_{B_1+B_2}\Bigl\{U_{42}(t)U_{42}^\dag(t)+
U_{43}(t)U_{43}^\dag(t)\Bigl\}.
\end{eqnarray}
Those associated with the initial state
$\frac{1}{\sqrt{2}}(|--\rangle\pm|++\rangle)$ read:
\begin{eqnarray}
\rho_{11}(t)&=&2^{-(2N+1)}\mathrm{tr}_{B_1+B_2}\Bigl\{U_{11}(t)U_{11}^\dag(t)+ U_{14}(t)U_{14}^\dag(t)\Bigl\},\\
\rho_{22}(t)&=&2^{-(2N+1)}\mathrm{tr}_{B_1+B_2}\Bigl\{U_{21}(t)U_{21}^\dag(t)+ U_{24}(t)U_{24}^\dag(t)\Bigl\},\\
\rho_{14}(t)&=&\pm2^{-(2N+1)}\mathrm{tr}_{B_1+B_2}\Bigl\{U_{11}(t)U_{44}^\dag(t)\Bigl\},\\
\rho_{33}(t)&=&2^{-(2N+1)}\mathrm{tr}_{B_1+B_2}\Bigl\{U_{31}(t)U_{31}^\dag(t)+ U_{34}(t)U_{34}^\dag(t)\Bigl\},\\
\rho_{44}(t)&=&2^{-(2N+1)}\mathrm{tr}_{B_1+B_2}\Bigl\{U_{41}(t)U_{41}^\dag(t)+
U_{44}(t)U_{44}^\dag(t)\Bigl\}.
\end{eqnarray}

\end{document}